\input ./style/arxiv-general.cfg
\documentclass[aoas,MSNbibl,nameyear,rotating,dvips]{arximspdf}
\makeatletter
   \@ifpackageloaded{graphicx}{}{\usepackage{graphicx}}
\makeatother
\usepackage{dcolumn}

\doi{10.1214/15-AOAS809}
\volume{9}
\issue{2}
\pubyear{2015}
\firstpage{572}
\lastpage{596}
\docsubty{FLA}

\makeatletter
\newcolumntype{d}[1]{D{.}{.}{#1}}
\def\mid{|}
\newcommand{\eqref}[1]{(\ref{#1})}
\newcommand{\E}{\mathbb{E}}
\newcommand{\btheta}{\bolds\theta}
\newcommand{\bgamma}{\bolds\gamma}
\newcommand{\bphi}{\bolds\phi}
\newcommand{\bbeta}{\bolds\beta}
\newcommand{\bomega}{\bolds\omega}
\newcommand{\balpha}{\bolds\alpha}
\newcommand{\bSigma}{\bolds\Sigma}
\newcommand{\Z}{\mathbf{Z}}
\newcommand{\W}{\mathbf{W}}
\newcommand{\CTMC}{U}
\newcommand{\gammaPone}{a}
\newcommand{\gammaPtwo}{b}
\makeatother

\begin{document}
\begin{frontmatter}

\title{Sex, lies and self-reported counts:
Bayesian mixture models for heaping in longitudinal count data
via birth--death processes\thanksref{TT1}}
\thankstext{TT1}{Supported by NIH Grants RR19895 and RR029676-01.}
\runtitle{Self-reported counts}

\begin{aug}
\author[A]{\fnms{Forrest W.}~\snm{Crawford}\thanksref{T1,M1}\ead[label=e1]{forrest.crawford@yale.edu}},
\author[B]{\fnms{Robert E.}~\snm{Weiss}\thanksref{T2,M2}\ead[label=e2]{robweiss@ucla.edu}}\\
\and
\author[C]{\fnms{Marc A.}~\snm{Suchard}\thanksref{T3,M2,M3}\ead[label=e3]{msuchard@ucla.edu}}
\runauthor{F.~W. Crawford, R.~E. Weiss and M.~A. Suchard}
\thankstext{T1}{Supported in part by NIH Grants T32GM008185 and KL2
TR000140.}
\thankstext{T2}{Supported in part by CHIPTS, NIH Grant 5P30MH058107 and
CFAR, NIH Grant AI 28697---CORE H.}
\thankstext{T3}{Supported in part by NSF Grants DMS 1264153 and IIS
1251151 and NIH
Grants
R01 HG006139 and R01 AI107034.}
\affiliation{Yale School of Public Health\thanksmark{M1},
UCLA Fielding School of Public Health\thanksmark{M2}  and
David Geffen School of Medicine at UCLA\thanksmark{M3}}
\address[A]{F.~W. Crawford\\
Department of Biostatistics\\
Yale School of Public Health\\
P.O. Box 208034\\
New Haven, Connecticut 06510\\
USA\\
\printead{e1}}
\address[B]{R.~E. Weiss\\
Department of Biostatistics\\
UCLA School of Public Health\\
Los Angeles, California 90095-1772\\
USA\\
\printead{e2}}
\address[C]{M.~A. Suchard\\
Departments of Biomathematics,\\
\quad Biostatistics and Human Genetics\\
6558 Gonda Building\\
695 Charles E. Young Drive\\
South Los Angeles, California 90095-1766\\
USA\\
\printead{e3}}
\end{aug}

%
\received{\smonth{5} \syear{2014}}
%
\revised{\smonth{2} \syear{2015}}

%
\begin{abstract}
Surveys often ask respondents to report nonnegative counts, but
respondents may misremember or round to a nearby multiple of 5 or 10.
This phenomenon is called heaping, and the error inherent in heaped
self-reported numbers can bias estimation. Heaped data may be collected
cross-sectionally or longitudinally and there may be covariates that
complicate the inferential task. Heaping is a well-known issue in many
survey settings, and inference for heaped data is an important
statistical problem. We propose a novel reporting distribution whose
underlying parameters are readily interpretable as rates of
misremembering and rounding. The process accommodates a variety of
heaping grids and allows for quasi-heaping to values nearly but not
equal to heaping multiples. We present a Bayesian hierarchical model
for longitudinal samples with covariates to infer both the unobserved
true distribution of counts and the parameters that control the heaping
process. Finally, we apply our methods to longitudinal self-reported
counts of sex partners in a study of high-risk behavior in HIV-positive
youth.
\end{abstract}

%
\end{frontmatter}

\section{Introduction}\label{sec1}

When survey respondents report numeric quantities, they often recall
those numbers with error. Respondents sometimes round up or down, for
example, to the nearest integer, decimal place or multiple of 5 or 10.
This kind of misreporting is called heaping, and when the probability
of heaping depends on the true value of the unheaped variable, the
mechanism is nonignorable [\citet{Heitjan1991Ignorability}]. Heaping is
a well-known problem in many survey settings, and robust inference for
heaped data remains an important problem in statistical inference
[\citet
{Heitjan1989Inference,Wang2008Modeling,Wright2003Mixture,Crockett2006Consequences,Schneeweiss2010Symmetric}].
Reporting errors are frequently observed for a variety of measurements,
including self-reported age [\citeauthor{Myers1954Accuracy}
(\citeyear{Myers1954Accuracy,Myers1976Instance}),
\citet{Stockwell1974Age}], height and
weight [\citet{Rowland1990Self,Schneeweiss2009Probabilistic}], elapsed
time [\citet{Huttenlocher1990Reports}] and household purchases [\citet
{Browning2003Asking}]. Respondents may be inclined to misreport when
the survey addresses topics that seem private, embarrassing or
culturally taboo [\citet{Schaeffer2000Asking}]. For example, there may
be significant misreporting in studies of drug use [\citet
{Klovdahl1994Social,Roberts2001Measures}], cigarette use [\citet
{Brown1998Reliability,Wang2008Modeling}] or number of sex acts or
sexual partners [\citet
{Westoff1974Coital,Golubjatnikov1983Homosexual,Wiederman1997Truth,Weinhardt1998Reliability,Fenton2001Measuring,Ghosh2009Assessing}].

Several authors have proposed approximations to correct estimates using
heaped data [\citet
{Sheppard1897Calculation,Schneeweiss2009Probabilistic,
Schneeweiss2010Symmetric,Schneeweiss2006Some,Tallis1967Approximate,
Lindley1950Grouping}].
Others have explored smoothing techniques for heaped data on the
grounds that smoothing may have the effect of ``spreading out'' grouped
responses [\citet{Hobson1976Properties,Singh1994Smoothed}].
\citet{Heitjan1989Inference} and \citeauthor{Heitjan1991Ignorability} (\citeyear
{Heitjan1990Inference,Heitjan1991Ignorability}) provide an important
unifying perspective on heaped and grouped data by introducing the
concept of coarsening, in which one observes only a subset of the
complete data sample space. Based on this paradigm, \citet
{Wang2008Modeling} formulate a model for heaped cigarette counts and
apply these ideas to study impact of a drug treatment on smoking.
\citet
{Jacobsen1995Coarsening} discuss extensions of the coarse data concept
to more general sample spaces than those considered by \citet
{Heitjan1991Ignorability}. \citet{Wright2003Mixture} model heaped
nuchal translucency measurements as samples from a mixture model and
propose a Gibbs sampling scheme to draw from the joint distribution of
the true counts and unknown rounding parameters. \citet
{Bar2012Accounting} model the age at which subjects quit smoking by
supposing that heaping takes place on a grid of multiples of 5 or 10.

\begin{figure}

\includegraphics{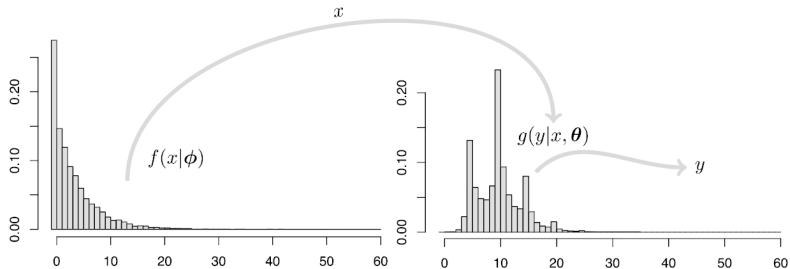}

\caption{Mixture model schematic for reported counts.
Each subject chooses their true count $x$ from the distribution
$f(x|\bphi)$, then reports the possibly different count
$y$ drawn from the distribution $g(y|x,\btheta)$.}\vspace*{-5pt}
\label{fig:schematic}
\end{figure}

Most attempts to disentangle heaped count responses from latent true
values can be understood as mixture models. To illustrate, suppose each
subject draws their latent true count $x$ from a distribution with mass
function $f(x|\bphi)$ on the nonnegative integers that depends on
parameters $\bphi$ and then reports a possibly different value $y$ from
a reporting distribution with mass function $g(y|x,\btheta)$ that
depends on the true count $x$ and parameters $\btheta$. Because the
reporting distribution $g$ depends on the latent true count $x$, the
heaping mechanism is nonignorable.
The likelihood contribution of an observed count $y$ is therefore
%
\begin{equation}
L(\btheta,\bphi;y) = \sum_{x=0}^\infty g(y|x,
\btheta) f(x|\bphi). \label{eq:lik}
\end{equation}
Figure~\ref{fig:schematic} shows a graphical representation of this
mixture model for heaped counts. The objects of inference are often the
true counts $x$ and the parameters $\bphi$ underlying the true count
distribution $f(x|\bphi)$.

Many approaches characterize the reporting mechanism as a choice
between reporting truthfully and misreporting at suspected heaping grid
points [e.g., \citet
{Wang2008Modeling,Wright2003Mixture,Wang2012Truth,Bar2012Accounting}].
The probability of reporting a particular heaped value depends on the
value of the latent true value: \citet{Wang2008Modeling} use a
proportional odds model for different heaping grids; \citet
{Bar2012Accounting} propose a multinomial distribution governing the
choice of different heaping rules; \citet{Mclain2014Semiparametric}
propose a semi-parametric model for heaping (digit preference) of
duration-time data in which subjects are equally likely to round up or down.
Most models for count data only allow exact heaping to the multiple of
5, 10 or 20 that is nearest to the latent true count, and the heaping
rule is the same for all subjects. However, limiting heaped responses
to the nearest grid point can produce inferences of true counts that
are unrealistically constrained. For example, if the reported count is
$y=35$ and the model only allows heaping to multiples of 5, then one
must infer $x\in\{33,\ldots,37\}$. Furthermore, established models do
not allow for misremembering as a function of the true count or
quasi-heaping to counts close to, but not equal to, the specified grid
values (e.g., a subject whose true count is 93 may report 101 or
99 instead of the heaped value 100).

In this paper, we relax several of these restrictive assumptions and
incorporate rigorous analysis of heaped data into a hierarchical
regression model. In Section~\ref{sec:bdp} we propose a novel reporting
distribution by imagining the true count $x$ as the starting point of a
continuous-time Markov chain on the nonnegative integers $\mathbb{N}$
known as a general birth--death process (BDP). The ending state of this
Markov chain after a specified epoch is the reported count $y$. Jumps
from integer state $k$ to $k+1$ or $k-1$ occur with instantaneous rates
$\lambda_k$ and $\mu_k$, respectively, with $\mu_0=0$ to keep the
process on $\mathbb{N}$. We specify $\lambda_k$ and $\mu_k$ so that the
process is attracted to nearby heaping grid points. Our BDP heaping
model characterizes an infinite family of reporting distributions
$g(y|x,\btheta)$ that is (1) indexed by the true count $x$; (2)
controlled by a small number of parameters $\btheta$ that are readily
interpretable; and (3) can be computed quickly to provide a reporting
likelihood. The model permits heaping to values beyond the nearest grid
point, provides for multiple heaping grids and continuous transitions
between them, allows misremembering and quasi-heaping, and accommodates
subject-specific heaping intensities. In Section~\ref{sec:model} we
outline a Bayesian hierarchical model for longitudinal counts and a
Metropolis-within-Gibbs scheme for drawing inference from the joint
posterior distribution of the unknown parameters. We are interested in
learning about the parameters $\bphi$ underlying the true counts, the
true counts $x$ themselves and the parameters $\theta$ that govern the
reporting/heaping process. Finally, in Section~\ref{sec:app} we
demonstrate our method on longitudinal self-reported counts of sexual
partners from a study of HIV-positive youth.


\section{Constructing the reporting distributions}

\label{sec:bdp}

Let $x$ be the true count for a subject and let $y$ be their reported
count. Let $g(y|x,\btheta)$ be the probability of reporting $y$, given
that their true count is $x$ under the parameter vector $\btheta$. To
parameterize $g(y|x,\btheta)$ to allow heaping, suppose $y$ represents
the state of an unbounded continuous-time Markov random walk, taking
values on $\mathbb{N}$, starting at $x$ and evolving for a finite
arbitrary time. We accomplish this task by defining the birth and death
rates $\lambda_k$ and $\mu_k$ of a general BDP in a novel way so that
the process is attracted to grid points on which we expect heaping to
occur. The transition probabilities of this process give rise to the
family of reporting distributions $g(y|x,\btheta)$. We extend the
proportional odds framework of \citet{Wang2008Modeling} to allow
heaping to different grid values depending on the magnitude of the
count. First we present background on general BDPs and show how to use
the transition probabilities of a general BDP to model heaping.

\subsection{General birth--death processes}
\label{sec:bdps}

A general BDP is a continuous-time Markov random walk on the
nonnegative integers $\mathbb{N}$ [\citet{Feller1971Introduction}]. Let
$\CTMC(t) \in\mathbb{N}$ be the location of the walk at time $t$.
Define the transition probability $P_{ab}(t) = \Pr(\CTMC(t)=b \mid
\CTMC
(0)=a)$ to be the probability that the process is in state $b$ at time~$t$,
given that it started at state $a$ at time $0$. A general BDP
obeys the Kolmogorov forward equations
%
\begin{equation}
\frac{\mathrm{d} P_{ab}(t)}{\mathrm{d} t} = \lambda _{b-1}P_{a,b-1}(t) +
\mu_{b+1}P_{a,b+1}(t) - (\lambda_b +
\mu_b)P_{ab}(t), \label{eq:odes}
\end{equation}
for all $a,b\in\mathbb{N}$, where $P_{ab}(0)=1$ if $a=b$,
$P_{ab}(0)=0$ if $a\neq b$, and $\mu_0=\lambda_{-1}=0$ to keep the BDP
on $\mathbb{N}$. In this setting, $t$ is arbitrary; for example,
halving $t$ and multiplying all birth and death rates by two does not
change the distribution of $\CTMC(t)|\CTMC(0)$. The forward equations
\eqref{eq:odes} form an infinite sequence of ordinary differential
equations describing the probability flow into and out of state $b$
within a small time interval $(t,t+\mathrm{d}t)$. \citet
{Karlin1957Differential} provide a detailed derivation of properties of
general BDPs. Unfortunately, it remains notoriously difficult to find
analytic expressions for the transition probabilities in almost all
general BDPs, and often one must resort to numerical techniques [\citet
{Novozhilov2006Biological,Renshaw2011Stochastic}]. Appendix~\ref
{app:laplace} gives an overview of the Laplace transform technique we
use to numerically compute the transition probabilities efficiently.

In our heaping parameterization, we model the true count $\CTMC(0)=x$
as the starting state of a BDP and $\CTMC(t)=y$ as the ending state. We
therefore set $t=1$ and define $g(y|x,\btheta) = P_{xy}(1)$ so that
$P_{xy}$ is a function of the unknown parameter vector $\btheta$, where
the $\{\lambda_k\}$ and $\{\mu_k\}$ are all functions of $\btheta$. We
emphasize that the time parameter $t$ is meaningless in this context,
because scaling $t$ by a constant and dividing the birth and death
rates by the same constant does not change the transition probabilities.


\subsection{Specifying the jumping rates \texorpdfstring{$\lambda_k$}{$lambda_k$} and \texorpdfstring{$\mu_k$}{$mu_k$}}
\label{sec:bdmodel}

Grunwald
et~al.
(\citeyear{Grunwald2011Statistical}) and \citet{Lee2011Using} model under-
and over-dispersion in count data using a simple linear BDP with
$\lambda_x=\mu_x=\lambda x$, but do not address heaping. In addition to
modeling dispersion, BDPs can be used to parameterize general families
of probability measures on $\mathbb{N}$ [\citet{Klar2010Zipf}]. In our
heaping model, we imagine errors in self-reported counts to come from
two sources: dispersion due to misremembering and heaping.
Misremembering adds variance by spreading reported counts around the
true count. Heaping results in preference for reporting certain counts,
for example, on a grid of values such as multiples of 5 or 10. We
specify both of these sources of misreporting error using a BDP with
jumping rates $\{\lambda_k\}$ and $\{\mu_k\}$ that are modeled as
functions of the finite-dimensional parameter vector $\btheta$.

To motivate development of our general BDP model for heaping, suppose
for now that heaping occurs at multiples of $5$. We wish to define a
random walk on $\mathbb{N}$ that is dispersed around its starting point
and attracted to multiples of 5, with this attraction increasing with
proximity to each multiple of 5. For example, if the true count is
$x=49$, then the reported count $y$ is more strongly attracted to $50$
than $45$, because 49 is closer to 50. Here, \emph{attraction} to a
given multiple means that the likelihood of the BDP moving toward that
multiple is greater than the likelihood of moving in the other
direction. Informally, we wish to assign birth and death rates such that
\begin{eqnarray}
\lambda_k &=& (\mbox{dispersion around } k) + (
\mbox{attraction to multiple of 5 above}),
\nonumber
\\[-8pt]
\\[-8pt]
\nonumber
\mu_k &=& (\mbox{dispersion around } k) + (\mbox{attraction to
multiple of 5 below}).
\nonumber
\end{eqnarray}
One way to quantify the strength of attraction to the multiple of 5
immediately above $k$ is $(k \mbox{ mod } 5)$.
Likewise, the attraction to the multiple of $5$ immediately below $k$
is $(-k \mbox{ mod } 5)$, which is equal to $5-(k \mbox{ mod } 5)$. 
In both directions, the closer $k$ is to the nearby multiple of 5, the
greater its attraction to it.

\begin{figure}

\includegraphics{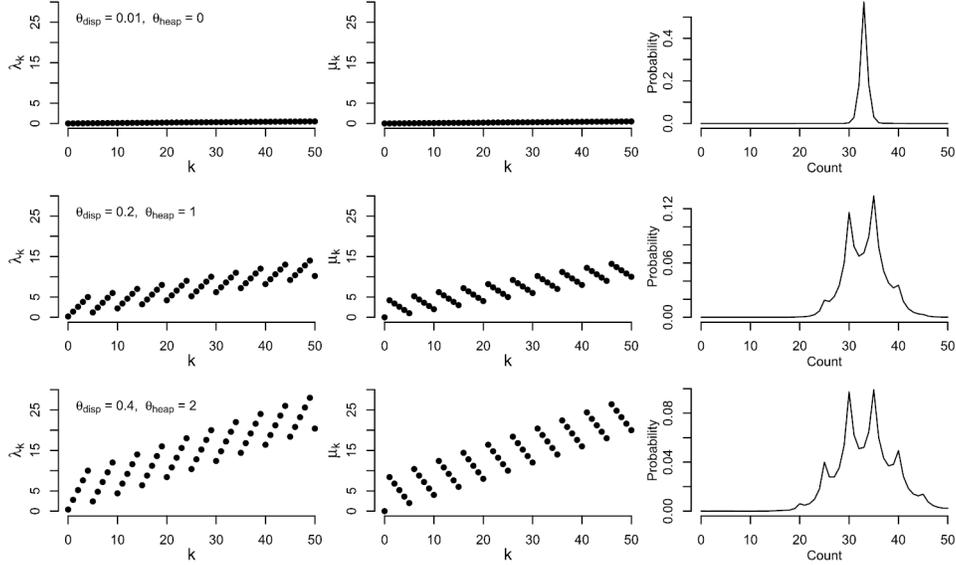}

\caption{Birth rates $\lambda_k$ (left), death rates $\mu_k$ (center)
and reporting probabilities for true count $x=33$ (right) in the
heaping model \protect\eqref{eq:bdrates5} for different values of the
dispersion parameter $\theta_\mathrm{disp}$ and heaping intensity
$\theta
_\mathrm{heap}$. Larger values of $\theta_\mathrm{disp}$ result in more
dispersion about the true count. Larger values of $\theta_\mathrm{heap}$
result in more heaping to nearby multiples of 5.}\vspace*{6pt}
\label{fig:heap5bdrates}
\end{figure}

Subjects whose true number of sex partners is greater than 100, for
example, may be less able to accurately recall this number than
subjects whose true count is less than 10. We therefore model
dispersion around the true count in the reported counts due to
misremembering as increasing the true count. Consider a general BDP
with jumping rates
%
\begin{eqnarray}
\label{eq:bdrates5} %
\lambda_k &=& \theta_\mathrm{disp}(1+k)
+ \theta_\mathrm{heap} (k \mbox{ mod } 5),
\nonumber
\\[-8pt]
\\[-8pt]
\nonumber
\mu_k &=& \theta_\mathrm{disp}k + \theta_\mathrm{heap}
(-k \mbox{ mod } 5) ,
\end{eqnarray}
where the $(1+k)$ in the birth rate arises because we wish to allow the
BDP to escape from zero with positive rate. In this formulation of the
birth and death rates, the dispersion parameter $\theta_\mathrm
{disp}\geq
0$ is the propensity to over- or under-report and $\theta_\mathrm
{heap}\geq0$ is the propensity of rounding up or down to multiples of
5. Figure~\ref{fig:heap5bdrates} shows the birth rates $\lambda_k$,
death rates $\mu_k$ and reporting probabilities with true count $x=33$
for this heaping model. The complexity of the reporting distributions
generated by the heaping model is evident in Figure~\ref
{fig:heap5bdrates}; the BDP tends toward multiples of 5 and the
magnitude of $\theta_\mathrm{heap}$ controls the severity of heaping. The
BDP heaping model exhibits subtler behavior than a dispersion
distribution with added mass at the heaping points.

\begin{figure}

\includegraphics{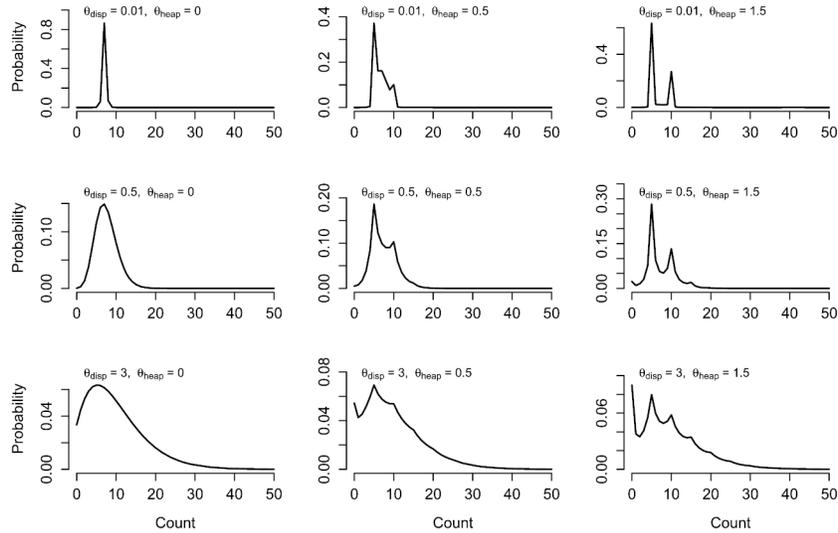}

\caption{Reporting probabilities for heaping at multiples of 5 with
true count $x=7$ using different values of the dispersion parameter
$\theta_\mathrm{disp}$ and the heaping parameter $\theta_\mathrm{heap}$.
Larger values of $\theta_\mathrm{disp}$ allow reports closer to zero;
when $\theta_\mathrm{heap}$ is positive, heaping occurs at zero,
providing a mechanism for zero-inflated reports.}
\label{fig:heap5params}
\end{figure}

Figure~\ref{fig:heap5params} shows reporting distributions for the true
count $x=7$. When $\theta_\mathrm{heap}=0$, the reporting distribution
only adds variance to the true count. As $\theta_\mathrm{heap}$ becomes
larger, the peaks in the reporting distribution at the heaping points
become more pronounced. When $\theta_\mathrm{heap}$ is large and
$\theta
_\mathrm{disp}$ is small, the reporting distribution is sharply peaked at
nearby multiples of 5 and the values between heaping points have little
probability mass.

\begin{figure}[b]

\includegraphics{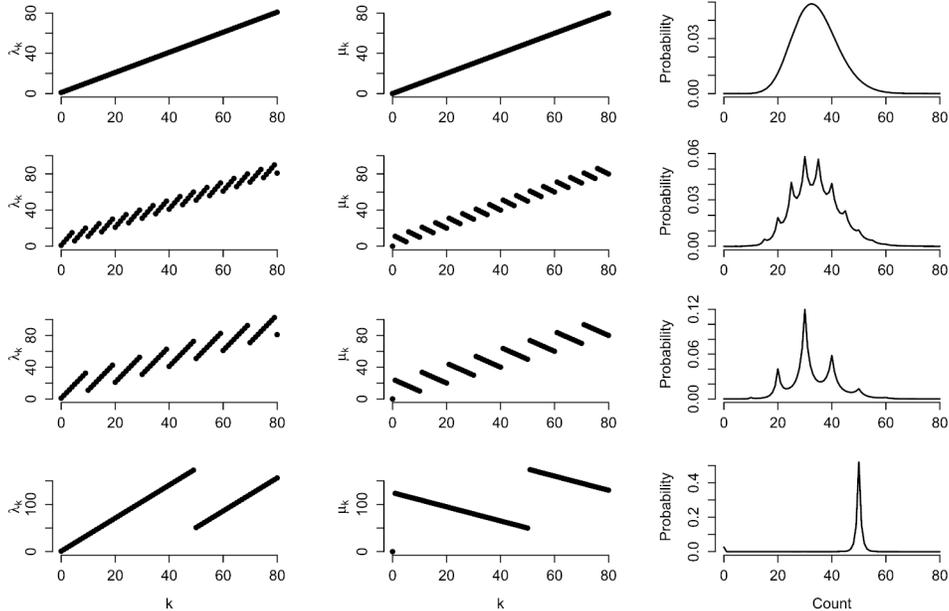}

\caption{Birth rates $\lambda_k$ (left), death rates $\mu_k$ (center)
and reporting probabilities (right) for different heaping grids with
true count $x=33$ and $\theta_\mathrm{disp}=1$. The first row shows the
reporting distribution for $\theta_\mathrm{heap}=0$. Subsequent rows show
the birth and death rates and reporting probabilities with $\theta
_\mathrm
{heap}=2.5$ with heaping at multiples of 5, 10 and 50. When heaping is
to multiples of 50 (bottom row), reporting is concentrated at $y=50$.}
\label{fig:rates}
\end{figure}

In general, suppose that heaping occurs at equally-spaced grid points
$mk$ where $m\in\mathbb{N}$ is the grid spacing; for example,\ $m$ could
be one of 5, 10, 20, 25 or 100. Analogous to \eqref{eq:bdrates5}, the
birth and death rates become
\begin{eqnarray}
\label{eq:bdrates2} %
\lambda_k &=& \theta_\mathrm{disp}(1+k)
+ \theta_\mathrm{heap} (k \mbox{ mod } m),
\nonumber
\\[-8pt]
\\[-8pt]
\nonumber
\mu_k &=& \theta_\mathrm{disp}k + \theta_\mathrm{heap}
(-k \mbox{ mod } m).
\nonumber
\end{eqnarray}
Figure~\ref{fig:rates} shows birth and death rates for several heaping
grid spacings $m$.

We can analytically characterize the properties of the reporting
distribution when $\theta_\mathrm{heap}$ is zero. Given the true count
$x$, the mean and variance of the reported count $y$ are
%
\begin{eqnarray}
\E[y\mid x] &=& x + \theta_\mathrm{disp} \quad\mbox{and}
\nonumber
\\[-8pt]
\\[-8pt]
\nonumber
\operatorname{Var}[y\mid x] &=& (2x+1)\theta_\mathrm{disp} + \theta_\mathrm{disp}^2.
\end{eqnarray}
Appendix \ref{app:X} provides a derivation of these expressions. It is
evident that both the mean and variance of $y|x$ increase linearly with
the true count $x$, consistent with our belief that the severity of
misremembering scales in proportion to the magnitude of the true count.

\subsection{Heaping regimes}

As true counts become larger, coarseness often increases; small counts
appear to be heaped at multiples of 5, then 10, and finally 50 or 100
for larger counts. Models such as \eqref{eq:bdrates5} that enforce
heaping to the same grid regardless of the magnitude of the count may
provide insufficient rounding behavior when the coarseness increases
with $x$. Consider $J$ distinct heaping grids and suppose $m_j$ is the
grid spacing for regime $j$, where $j=1,\ldots,J$. Let $v_j(x)$ be the
intensity of regime $j$ as a function of the true count $x$. Regime
$0$, with intensity $v_0(x)$, is the probability of accurately
reporting the true count. Regime $j$, with intensity $v_j(x)$,
corresponds to heaping at multiples of $m_j$. We follow \citet
{Wang2008Modeling} to develop a proportional odds model for smooth
transitions between heaping grids.

\begin{figure}[b]

\includegraphics{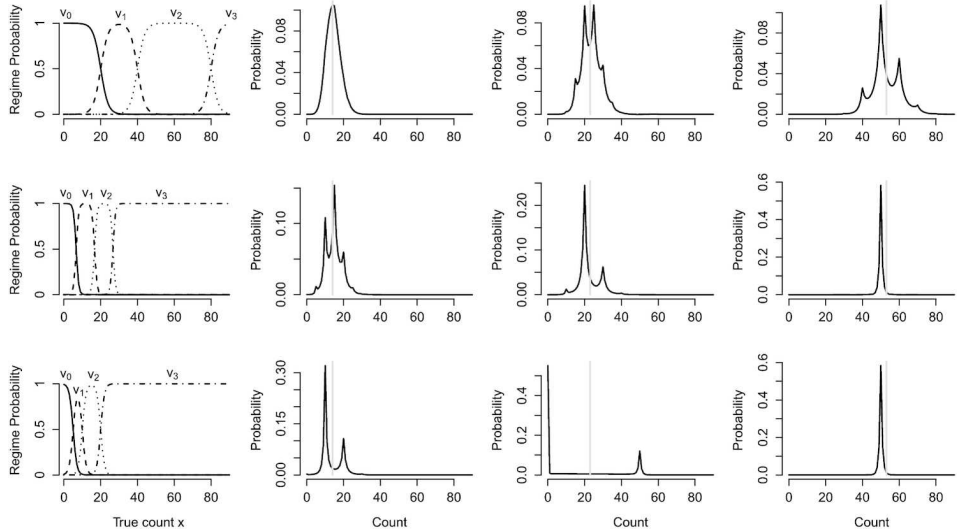}

\caption{Heaping regimes. Each row shows a different heaping regime
model with reporting probabilities for $\theta_\mathrm{disp}=0.5$ and
$\theta_\mathrm{heap}=1.5$. A gray line denotes the true counts
$x=14,23,53$. In the first row, the regime intensities are shown with
regime parameters $\bgamma=(0.5,-10,-20,-40)$. For $x=14$, the
reporting distribution is dominated by regime 0, which specifies no
heaping. For $x=23$, the reporting distribution is dominated by regime
1, so rounding to nearby multiples of 5 is evident. At $x=53$, regime 2
is dominant, and the reporting distribution is peaked at multiples of
10. In the second row, $\bgamma=(1.5,-10,-25,-40)$, and the reporting
distribution for $x=53$ is dominated by regime 3, so the model exhibits
heaping to multiples of 50. In the third row, $\bgamma=(1,-5,-10,-20)$.}
\label{fig:regimes}
\end{figure}

Define birth and death rates
%
\begin{eqnarray}
\label{eq:bdratesmult} %
\lambda_k &=& \theta_\mathrm{disp}(1+k)
+ \theta_\mathrm{heap}\sum_{j=1}^J
v_j(x) (k \mbox{ mod } m_j ),
\nonumber
\\[-8pt]
\\[-8pt]
\nonumber
\mu_k &= &\theta_\mathrm{disp}k +\theta_\mathrm{heap}
\sum_{j=1}^J v_j(x) (-k \mbox{
mod } m_j),
\end{eqnarray}
where the heaping regime probabilities are
%
\begin{eqnarray}
\label{eq:vs} %
v_0(x) &=& \bigl( 1 + e^{\gamma_1 + \gamma_0 x}
\bigr)^{-1},\nonumber
\\
v_1(x) &=& \bigl( 1 + e^{\gamma_2 + \gamma_0 x} \bigr)^{-1} - \bigl(
1 + e^{\gamma_1 + \gamma_0 x} \bigr)^{-1},
\\
v_2(x) &=& \bigl( 1 + e^{\gamma_3 + \gamma_0 x} \bigr)^{-1} - \bigl(
1 + e^{\gamma_2 + \gamma_0 x} \bigr)^{-1},\nonumber
\\
&\vdots&\nonumber
\\
v_J(x) &=& 1 - \bigl( 1 + e^{\gamma_J + \gamma_0 x} \bigr)^{-1},
\nonumber
\end{eqnarray}
and we restrict the regime transition parameters $\gamma_0>0$ and
$\gamma_1>\gamma_2>\cdots>\gamma_J$. We have, by construction,
%
\begin{equation}
\sum_{j=1}^J v_j(x) = 1 ,
\end{equation}
for every $x\in\mathbb{N}$. In this proportional odds model, $\gamma_0$
determines the transition rate between regimes and $\gamma_j/\gamma_0$
controls the midpoint of the transition between regimes $j-1$ and $j$.
Figure~\ref{fig:regimes} shows the heaping regime model defined above.
Each row shows a different heaping regime model and reporting
distribution $g(y|x,\btheta,\bgamma)$, where $\bgamma= (\gamma_0,
\ldots, \gamma_J)$ for $x=14,23,53$ and $\btheta=(0.5,1.5)$.

\subsection{Justification for the BDP heaping model}

We formulate the heaping model as a continuous-time Markov process for
three reasons: mathematical convenience, diversity of reporting
distributions, and parsimony in parameterization. First, the theory of
general BDPs is well developed and efficient methods now exist for
computing transition probabilities for any specification of the birth
and death rates [\citet{Crawford2012Transition}]. The heaping
probability mass function $g(y|x)$ is automatically normalized to
integrate to one (since it is the likelihood of a Markov process), so
the mixture model \eqref{eq:lik} is always well defined. Second, the
model described in \eqref{eq:bdratesmult} and \eqref{eq:vs} exhibits a
great diversity in reporting distributions, from no heaping to always
heaping, under a wide variety of magnitude-based regimes (see
Figures~\ref{fig:heap5bdrates}--\ref{fig:regimes}, e.g.). Third, the general BDP achieves this complex
behavior using only two parameters for the heaping process and four in
the regimes specification. Additionally, the specification of heaping
regimes via \eqref{eq:bdratesmult} and \eqref{eq:vs} results in an
appealing property: the reporting distribution can by highly
asymmetrical when the true count is subject to two heaping regimes. For
example, the third row of Figure~\ref{fig:regimes} shows how the true
count $x=14$ can be pulled toward 10 and 20 with very different probabilities.


\section{A hierarchical model for longitudinal counts}
\label{sec:model}

We describe a generalized linear mixed model (GLMM) for longitudinal
counts. Label subjects $i=1,\ldots,N$, with each subject's true count
$X_{it}$ and self-reported count $Y_{it}$ at real calendar timepoints
$t_{ij}$ for $j=1,\ldots,n_i$. We record $d$-dimensional covariates
$\W
_{it}$ and $c$-dimensional $\Z_{it}$ for each subject at each
timepoint. Consider the following hierarchical model:
%
\begin{eqnarray}
X_{it} &\sim&\operatorname{Poisson}(\eta_{it}), \label{eq:model:poisson}
\\
\log\eta_{it} &= &\W_{it} \balpha+ \Z_{it}
\bbeta_i \label{eq:model:eta}
\end{eqnarray}
and
\begin{equation}
\bbeta_i \sim\operatorname{Normal}(\mathbf{0},
\bSigma_\beta), \label{eq:model:beta}
\end{equation}
where the vector of regression coefficients $\balpha$ is $d\times1$, the
subject-specific random effect $\bbeta_i$ is $c\times1$ with the
covariance matrix
$\bSigma_\beta$ is $c\times c$, and $\eta_{it}$ is the
subject-timepoint-specific mean of the outcome distribution in the GLMM.

A model without heaping arises when we set $Y_{it}=X_{it}$ for all $i$
and $t$. To incorporate heaping, let
%
\begin{equation}
Y_{it} \sim\operatorname{BDP}(X_{it},\btheta,\bgamma).
\end{equation}
We allow the BDP heaping model to have a separate heaping intensity
parameter $\theta_{\mathrm{heap},i}$ for each subject. If $X_{it}=x$, the
birth and death rates for subject $i$ are
%
\begin{eqnarray}
\label{eq:clearbdrates} %
\lambda_k &=& \theta_\mathrm{disp}(1+k)
+ \theta_{\mathrm{heap},i} \sum_{j=1}^3
v_j(x) (k \mbox{ mod } m_j) \quad\mbox{and}
\nonumber
\\[-8pt]
\\[-8pt]
\nonumber
\mu_k &=& \theta_\mathrm{disp}k + \theta_{\mathrm{heap},i} \sum
_{j=1}^3 v_j(x)
\bigl(m_j - \bigl((k-1) \mbox{ mod } m_j \bigr) \bigr),
\end{eqnarray}
where $m_1=5$, $m_2=10$, $m_3=50$, and $v_1(x)$, $v_2(x)$, and $v_3(x)$
are defined above in \eqref{eq:vs}. The subject-specific heaping
intensity is
%
\begin{equation}
\log\theta_{\mathrm{heap},i} = \mathbf{H}_{i} \bomega+
\xi_i , \label{eq:xi}
\end{equation}
where $\mathbf{H}_{i}$ is a heaping covariate vector for subject $i$,
$\bomega$ is an unknown parameter vector of corresponding dimension,
and $\xi_i$ is a subject-specific random effect, with distribution
%
\begin{equation}
\xi_i \sim\operatorname{Normal}(\mathbf{0},\sigma_\xi).
\end{equation}
To complete our Bayesian hierarchical model for longitudinal studies,
we specify conditionally conjugate prior distributions for $\balpha$
and $\bSigma_\beta$: 
%
\begin{eqnarray}
\label{eq:priors} %
\balpha&\sim&\operatorname{Normal}(\mathbf{0},
\mathbf{V}_\alpha),\nonumber
\\
\theta_\mathrm{disp} &\sim&\operatorname{Inverse\mbox{-}Gamma}(\gammaPone, \gammaPtwo),
\nonumber\\
\bomega&\sim&\operatorname{Normal}(\mathbf{0},\bSigma_\omega),
\\
\bgamma&\sim&\operatorname{Normal}(\mathbf{0}, \mathbf{V}_\gamma)\qquad \mbox{subject
to } \gamma_0<\cdots<\gamma_J\quad \mbox{and}\nonumber
\\
\bSigma_\beta&\sim&\operatorname{Inverse\mbox{-}Wishart}(A_\beta,
\mathbf{m}_\beta),\nonumber 
\end{eqnarray}
%
where $\mathbf{V}_\alpha$, $\gammaPone$, $\gammaPtwo$, $\mathbf
{V}_\gamma$, $A_\beta$ and $\mathbf{m}_\beta$ are fixed hyperparameters
of corresponding dimension that we specify in Section~\ref{sec:app}.

Finally, we fit an alternative model of \citet{Wang2008Modeling} in
which responses not equal to a heaping point are assumed to be reported
accurately. The model for the latent counts $X_{it}$ is identical to
\eqref{eq:model:poisson}--\eqref{eq:model:beta}, but the heaping
distribution is different. If $x$ is the true count, then $y$ is
reported as
%
\begin{equation}
\label{eq:wangcases} y = %
\cases{ x, &\quad $\mbox{with probability }
v_0(x),$ \vspace*{2pt}
\cr
\mbox{nearest multiple of 5}, &\quad $\mbox{with probability } v_1(x),$ \vspace*{2pt}
\cr
\mbox{nearest multiple
of 10}, &\quad $ \mbox{with probability }v_2(x),$ \vspace*{2pt}
\cr
\mbox{nearest multiple of 50}, & \quad$\mbox{with probability }v_3(x)$.}
\end{equation}
Once the heaping regime in \eqref{eq:wangcases} has been determined,
the reported count $y$ arises deterministically.

\subsection{Posterior inference}
\label{sec:sampling}

We estimate the joint posterior distribution with Markov chain Monte
Carlo (MCMC). We describe standard
Gibbs and Metropolis--Hastings
samplers for the full conditional distributions of $\balpha$, $\bbeta=
(\bbeta_1,\ldots, \bbeta_N)$, $\btheta$, $\bgamma$ and $\bSigma
_\beta$
in the supplemental material [\citet{Crawford2015HeapingSupplement}].
Sampling from the conditional posterior distribution of the true counts
is more challenging because of the lack of conjugacy between $\Pr
(X_{it}|\Z_{it},\W_{it},\balpha,\bbeta_i)$ and
$g(Y_{it}|X_{it},\btheta
)$. Fortunately, the discrete nature of count data makes some
simplifications possible. The conditional distribution of the
unobserved true count $X_{it}$ is 
%
\begin{eqnarray}
&&\Pr(X_{it} \mid Y_{it},\Z_{it},\W_{it},
\mathbf{H}_i,\btheta,\balpha,\bbeta_i)
\nonumber
\\[-8pt]
\\[-8pt]
\nonumber
&&\qquad\propto
g(Y_{it}\mid X_{it},\btheta) \Pr(X_{it} \mid\Z
_{it},\W_{it},\mathbf{H}_i,\balpha,
\bbeta_i).
\end{eqnarray}
It is computationally costly to evaluate $g(y|x,\btheta)$ hundreds of
times to construct the distribution of $X_{it}$. In the \hyperref
[app]{Appendix} we
present a method for approximating this density by a discretized normal
distribution derived from the dynamics of the BDP with $\theta_\mathrm
{heap}=0$, allowing efficient sampling. We then employ a
Metropolis--Hastings accept/reject step to sample from the correct posterior.


\section{Simulation study}

To validate the proposed heaping model and the associated Bayesian
inference framework, we simulate data under a simplification of the
hierarchical model described in Section~\ref{sec:model}:
%
\begin{eqnarray}
\label{eq:sim} %
Y_{it} &\sim&\operatorname{BDP}(X_{it},
\btheta, \bgamma) ,\nonumber
\\
X_{it} &\sim&\operatorname{Poisson}(\eta_{it}) ,
\nonumber
\\[-8pt]
\\[-8pt]
\nonumber
\log\eta_{it} &= &\alpha+ \beta_i \quad \mbox{and}
\\
\beta_i &\sim&\operatorname{Normal} \bigl(0,\sigma^2_\beta
\bigr) ,\nonumber %
\end{eqnarray}
for subjects $i=1,\ldots,n$ and repeated measures $t=1,\ldots,5$, with
$\alpha$ and $\beta_i$ scalars. The heaping parameter $\theta_{\mathrm
{heap},i}=\theta_\mathrm{heap}$ is constant for every subject. Setting
$\alpha=2$, $\sigma^2_\beta=1.21$, $\bgamma=(0.5,-5,-10,-20)$, and
$\theta_\mathrm{disp}=0.5$ and $\theta_\mathrm{heap}=2$ yields observed
counts qualitatively similar to those we observe in the application
section below. From this model, we simulate data sets with $N=100$, 250
and 500 total observations from $n=N/5$ subjects. Using $100$
replicates, Table~\ref{tab:sim} reports true parameter values, average
posterior means, average posterior variances and mean squared error
(MSE) for each data set. Standard deviations are given in parentheses.
As expected, simulations with larger $N$ give, in general, more
accurate parameter estimates, with posterior variance and MSE
decreasing with~$N$. Posterior mean estimates of the heaping regimes
parameters $\gamma_2$ and $\gamma_3$ are close to their true
values, but their MSE does not appear to decrease monotonically with~$N$.
The regime parameters may be only weakly identified in data sets
with few large reported counts. Since these parameters control the
midpoints of transitions between heaping regimes, they may be highly
variable unless many counts fall near these transitions. In addition to
larger $N$, it may be necessary to observe a greater proportion of
heaped counts near regime transitions in order to achieve a substantial
reduction in posterior variance for $\gamma_2$ and $\gamma_3$.

\begin{sidewaystable}
\tablewidth=\textwidth
\tabcolsep=0pt
\caption{Summary of estimated parameters from 100 simulated datasets of
size $N=100$, $250$ and $500$ under the heaping model given by \protect
\eqref
{eq:sim}. Averages of the posterior means, averages of the posterior
variances and mean squared errors are shown with standard deviations in
parentheses}\label{tab:sim}
\begin{tabular*}{\textwidth}{@{\extracolsep{\fill}}ld{3.2}d{3.10}ccd{3.10}ccd{3.10}cc@{}}
\hline
&& \multicolumn{3}{c}{$\bolds{N=100}$} & \multicolumn{3}{c}{$\bolds{N=250}$} &
\multicolumn{3}{c@{}}{$\bolds{N=500}$} \\ [-6pt]
&& \multicolumn{3}{c}{\hrulefill} & \multicolumn{3}{c}{\hrulefill} &
\multicolumn{3}{c@{}}{\hrulefill} \\
& \textbf{True} & \multicolumn{1}{c}{\textbf{Mean}} & \multicolumn{1}{c}{\textbf{Var}} &
\multicolumn{1}{c}{\textbf{MSE}} &
\multicolumn{1}{c}{\textbf{Mean}} & \multicolumn{1}{c}{\textbf{Var}} & \multicolumn{1}{c}{\textbf{MSE}} &
\multicolumn
{1}{c}{\textbf{Mean}} & \multicolumn{1}{c}{\textbf{Var}} & \multicolumn{1}{c@{}}{\textbf{MSE}} \\
\hline
$\alpha$ & 2.00 & 1.991\ (0.28) & 0.059 (0.03) & 0.078 & 1.970\ (0.14) &
0.026 (0.01) & 0.021 & 2.030\ (0.12) & 0.013 (0.00) & 0.014 \\
$\sigma^2_\beta$ & 1.21 & 1.368\ (0.32) & 0.210 (0.12) & 0.123 & 1.270\
(0.22) & 0.081 (0.03) & 0.052 & 1.211\ (0.16) & 0.037 (0.01) & 0.027 \\
$\theta_\mathrm{disp}$ & 0.50 & 0.516\ (0.17) & 0.026 (0.01) & 0.030 &
0.508\ (0.09) & 0.011 (0.00) & 0.008 & 0.492\ (0.08) & 0.006 (0.00) &
0.007 \\
$\theta_\mathrm{heap}$ & 2.00 & 2.013\ (1.11) & 0.572 (0.68) & 1.220 &
2.288\ (0.89) & 0.615 (0.90) & 0.858 & 2.157\ (0.71) & 0.368 (0.61) &
0.527 \\
$\gamma_0$ & 0.50 & 0.494\ (0.08) & 0.004 (0.01) & 0.007 & 0.497\ (0.07)
& 0.003 (0.00) & 0.005 & 0.492\ (0.06) & 0.003 (0.00) & 0.004 \\
$\gamma_1$ & -5.00 & -5.022\ (1.37) & 0.840 (0.56) & 1.867 & -5.204\
(0.92) & 0.657 (0.48) & 0.881 & -5.231\ (0.86) & 0.616 (0.46) & 0.790 \\
$\gamma_2$ & -10.00 & -9.677\ (1.70) & 1.617 (1.31) & 2.949 & -9.916\
(1.41) & 1.516 (0.98) & 1.985 & -10.282\ (1.50) & 1.418 (0.91) & 2.290
\\
$\gamma_3$ & -20.00 & -19.603\ (2.21) & 2.343 (2.05) & 4.969 & -19.388\
(2.17) & 3.126 (2.58) & 5.050 & -19.351\ (2.26) & 3.250 (2.09) & 5.486
\\
\hline
\end{tabular*}
\end{sidewaystable}

\section{Application to self-reported counts of sex partners}
\label{sec:app}

\begin{figure}

\includegraphics{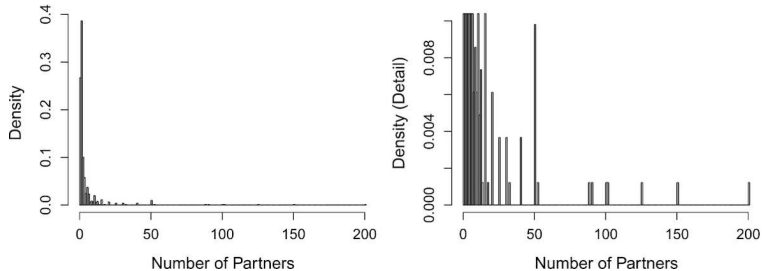}

\caption{Summary of self-reported counts of sex partners. At left, the
histogram shows the aggregate reported number of partners in the
previous three months, for all subjects, at all timepoints. At right is
the same histogram with the vertical axis limited to $(0,0.01)$ to show
greater detail. There is an apparent preference for reporting counts in
multiples of 5, 10 and 50.}
\label{fig:summary}
\end{figure}

To illustrate the effectiveness of our mixture model and general BDP
characterization of the reporting distributions $g(y|x,\btheta)$, we
analyze a survey of HIV-positive youth regarding their sexual behavior
from the Choosing Life: Empowerment, Action Results (CLEAR)
longitudinal three-arm randomized intervention study designed to reduce
HIV transmission and improve quality of life [\citet
{Rotheram-Borus2001Efficacy}]. Respondents (175, interviewed between 2
and 5 times for 816 total observations) report the number of unique sex
partners they had during the previous three months. Figure~\ref
{fig:summary} summarizes the reported counts. There are several
striking features of the reported counts: (1) a fair proportion ($27\%
$) of the counts are zero; (2) the histogram shows peaks at integer
multiples of 10; and (3) a few counts are very large.

We let $\W_{it}$ in \eqref{eq:model:eta} be an $8\times1$ vector of
covariates for subject $i$ at time $t$ by including subject baseline
age, gender (1 for male, 0 for female), an indicator for men who have
sex with men (MSM), an indicator for injection drug use, time since
baseline interview, an indicator for post-baseline educational
intervention and an indicator for use of methamphetamine or other
stimulant drugs. Time since baseline interview, use of drugs and
post-baseline intervention, depend on the timepoint $t$. To facilitate
comparison of estimated effects, subject age and time since baseline
interview were standardized by subtracting the mean and dividing by the
standard deviation. We let $\Z_{it} = 1$, making $\bbeta_i$ a scalar;
this provides a subject-specific random intercept. We fit two
subject-specific heaping models. In the first, we let $\mathbf
{H}_{i}=1$ so that $\theta_{\mathrm{heap},i}$ is a subject-specific
random intercept. In the second, $\mathbf{H}_{i} = (1,\mathrm{gender})$.
Based on the histogram of aggregate counts in Figure~\ref{fig:summary},
we use the BDP rate model in equation \eqref{eq:bdratesmult} with $J=3$
regimes corresponding to heaping at grid points at multiples of 5, 10
or 50.

We assign hyperparameters as follows: for the fixed effects $\balpha$,
$\balpha_0=\mathbf{0}$ and $\bSigma_\alpha=10 \mathbf{I}$ where
$\mathbf
{I}$ is the identity matrix; for the heaping parameters $\btheta$,
$\gammaPone= 0.001$ and $\gammaPtwo= 0.001$, such that each has a
prior expectation of 1 and variance 1000; for $\bgamma$, $\sigma
^2_\gamma=100$. Since the subject-specific random effects $\bbeta_i$
are scalars, $\bbeta_i$ has inverse gamma distribution with parameters
$A_\beta=4$ and $\mathbf{m}_\beta=5$.


\subsection{Results}

To evaluate the usefulness of our heaping distributions and to compare
to previous approaches, we fit six hierarchical Bayesian models: (1)~Poisson mixed effects (PME) with $X_{it}=Y_{it}$ and no heaping; (2)
the model of \citet{Wang2008Modeling} (WH08) as defined by \eqref
{eq:wangcases}; (3) BDP with dispersion and no heaping; (4) BDP model
with dispersion and global heaping parameter $\theta_\mathrm{heap}$; (5)
BDP model with subject-specific heaping intensity; and (6) BDP model
with subject-specific heaping intensity and a fixed effect controlling
heaping propensity for male and female subjects. In each case, the
model for the underlying true count is identical to \eqref
{eq:model:poisson}--\eqref{eq:model:beta}. The priors on equivalent
parameters are also the same for all models.

\begin{sidewaystable}
\tabcolsep=0pt
\tablewidth=\textwidth
\caption{Parameter estimates, intervals, and goodness-of-fit measures
of the CLEAR data. We fit six models, each using the basic Bayesian
Poisson regression setup \protect\eqref{eq:model:poisson} for the
true counts.
In the model without heaping, the reported counts are assumed to be
equal to true counts. In the dispersion-only model, the BDP allows
misremembering but not heaping. The \citet{Wang2008Modeling} model
involves deterministic heaping under different regimes \protect\eqref
{eq:wangcases}. The BDP heaping model has global dispersion and heaping
parameters, the subject-specific BDP heaping model allows
subject-specific effects \protect\eqref{eq:xi}, and the
subject-specific model
with covariates includes a fixed effect for the influence of gender on
heaping behavior. Parameter estimates (posterior means) and 95\%
posterior quantiles are shown for each parameter. The fixed effects are
age, gender, men who have sex with men (MSM), injection drug user,
intervention, stimulant use and trading sex. The random intercept
variance $\sigma^2_\beta$ is also shown. The heaping parameters
$\theta
_\mathrm{disp}$ and $\theta_\mathrm{heap}$ control dispersion and heaping
for the BDP models. The heaping regime parameters $\gamma_0$, $\gamma
_1$, $\gamma_2$ and $\gamma_3$ are shown for the heaping models. The
heaping random intercept variance $\sigma^2_\xi$ and the
gender-specific heaping fixed effect $\omega$ are also shown. Finally,
we provide two measures of goodness of fit for each model: deviance
information criterion (DIC) and the sum of squared mean prediction
errors, and the sum of squared prediction errors (SSPE)}
\label{tab:results}
{\fontsize{8}{10}\selectfont
\begin{tabular*}{\textwidth}{@{\extracolsep{\fill}}ld{2.2}cd{2.2}cd{2.2}cd{2.2}cd{2.2}cd{2.2}c@{}}
\hline
& && &&  &&  && \multicolumn{2}{c}{\textbf{Subject-specific}} &
\multicolumn{2}{c@{}}{\textbf{Subject-specific}} \\
& \multicolumn{2}{c}{\textbf{No heaping}} & \multicolumn{2}{c}{\textbf{WH08}} &
\multicolumn{2}{c}{\textbf{Dispersion-only}} & \multicolumn{2}{c}{\textbf{Heaping}} &
\multicolumn{2}{c}{\textbf{heaping}} &
\multicolumn{2}{c@{}}{\textbf{heaping${}\bolds{+}{}$gender}} \\
\hline
Age & -0.11 & $(-0.27,0.08)$ & -0.07 & $(-0.25,0.1)$ & -0.15 &
$(-0.56,0.25)$ & -0.20 & $(-0.58,0.14)$ & -0.12 & $(-0.55,0.23)$ & -0.14 &
$(-0.43,0.22)$ \\
Male & -0.26 & $(-0.78,0.25)$ & -0.24 & $(-0.74,0.28)$ & -1.48 &
$(-2.77,-0.27)$ & -0.85 & $(-1.81,0.12)$ & -1.01 & $(-2.16,-0.01)$ & -0.98
& $(-2.05,-0.02)$ \\
MSM & 0.82 & $(0.33,1.32)$ & 0.81 & $(0.3,1.32)$ & 0.57 & $(-0.59,1.75)$
& 0.89 & $(-0.06,1.85)$ & 0.99 & $(0.03,1.99)$ & 0.92 & $(-0.06,1.95)$ \\
Inject & -0.37 & $(-0.88,0.11)$ & -0.29 & $(-0.72,0.18)$ & -0.38 &
$(-1.45,0.65)$ & -0.29 & $(-1.2,0.56)$ & -0.35 & $(-1.3,0.55)$ & -0.38 &
$(-1.54,0.44)$ \\
Time & -0.89 & $(-1.06,-0.72)$ & -0.85 & $(-1.03,-0.66)$ & -1.72 &
$(-2.27,-1.18)$ & -1.02 & $(-1.46,-0.6)$ & -1.09 & $(-1.51,-0.67)$ & -1.06
& $(-1.5,-0.61)$\\
Intv & -0.24 & $(-0.57,0.05)$ & -0.18 & $(-0.5,0.1)$ & -1.29 &
$(-2.05,-0.6)$ & -1.07 & $(-1.76,-0.45)$ & -1.09 & $(-1.85,-0.32)$ & -1.16
& $(-2.06,-0.47)$ \\
Stim & 1.00 & $(0.88,1.12)$ & 0.97 & $(0.84,1.1)$ & 1.51 & $(1.14,1.88)$
& 1.09 & $(0.82,1.39)$ & 1.15 & $(0.83,1.47)$ & 1.05 & $(0.77,1.36)$ \\
Trade & 1.32 & $(1.2,1.45)$ & 1.21 & $(1.08,1.35)$ & 2.49 & $(1.98,3)$ &
1.81 & $(1.41,2.21)$ & 1.79 & $(1.44,2.15)$ & 2.00 & $(1.65,2.34)$ \\
$\sigma^2_\beta$ & 1.15 & $(0.88,1.48)$ & 1.07 & $(0.82,1.36)$ & 3.63 &
$(2.2,5.66)$ & 2.77 & $(1.75,4.47)$ & 2.93 & $(1.79,4.81)$ & 2.93 &
$(1.93,4.45)$ \\[3pt]
$\theta_\mathrm{disp}$ &&&&&  1.57 & $(1.4,1.75)$ & 1.04 & $(0.86,1.22)$
& 1.08 & $(0.9,1.27)$ & 1.06 & $(0.91,1.24)$ \\
$\theta_\mathrm{heap}$ &&&&& & & 0.82 & $(0.59,1.12)$ && && \\
$\gamma_0$ &&& 0.07 & $(0.05,0.11)$ &&&  0.42 & $(0.26,0.84)$ & 0.29
& $(0.21,0.4)$ & 0.45 & $(0.28,0.79)$ \\
$\gamma_1$ &&& -2.37 & $(-2.86,-1.95)$ &&&  -4.51 & $(-6.09,-3.46)$ &
-4.66 & $(-5.78,-3.68)$ & -5.50 & $(-8.43,-4.16)$ \\
$\gamma_2$ &&& -2.90 & $(-3.47,-2.42)$ &&&  -5.44 & $(-8.75,-3.95)$ &
-5.40 & $(-6.87,-4.21)$ & -7.23 & $(-10.11,-5)$ \\
$\gamma_3$ &&& -4.07 & $(-4.9,-3.39)$ &&&  -6.81 & $(-12.47,-4.75)$ &
-6.22 & $(-7.68,-5.1)$ & -8.40 & $(-11.55,-6.36)$ \\
$\sigma^2_\xi$ &&&&& & && & 0.74 & $(0.61,0.98)$ & 0.94 &
$(0.87,1)$ \\
$\omega$ &&&&& & && & && -0.03 & $(-0.69,0.54)$ \\[3pt]
DIC & \multicolumn{2}{c}{4585} & \multicolumn{2}{c}{524} &
\multicolumn{2}{c}{3329} & \multicolumn{2}{c}{3214} & \multicolumn
{2}{c}{3195} & \multicolumn{2}{c}{3175} \\
SSPE & \multicolumn{2}{c}{47,773} & \multicolumn{2}{c}{55,078} &
\multicolumn{2}{c}{28,005}& \multicolumn{2}{c}{25,371} &
\multicolumn
{2}{c}{25,336} & \multicolumn{2}{c}{24,364} \\
\hline
\end{tabular*}}
\end{sidewaystable}

\begin{figure}

\includegraphics{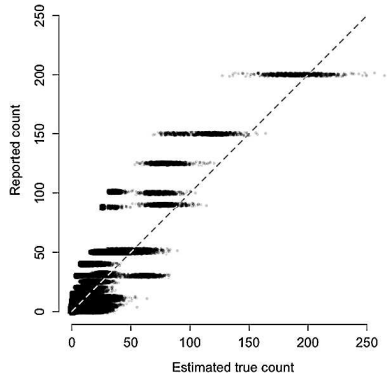}

\caption{Posterior samples of true counts on the horizontal axis versus
reported counts on the vertical axis for the CLEAR data under the BDP
heaping model. The points have been slightly jittered to show the
density of posterior samples. A gray dashed line is shown on the
diagonal.}
\label{fig:xyplot}
\end{figure}

Table~\ref{tab:results} shows posterior summaries for each model. The
first eight rows are regression coefficients for the fixed effects
$\balpha$. Estimates of fixed effects in the WH08 model are similar to
those found in the PME model without heaping. In general, fixed effects
estimates all have larger variance in the heaping models because the
BDP reporting distribution induces over-dispersion. Use of stimulants
is positively associated with increased true count. While the
intervention is not significantly associated with decreased reported
counts in the model without heaping and in the \citet{Wang2008Modeling}
model, the intervention has a clear association with reduced true
counts in the BDP heaping models. This result suggests that heaping in
reported counts may obscure important associations between covariates
and count outcomes. Figure~\ref{fig:xyplot} plots the posterior
distribution of true counts $X_{it}$ versus their corresponding
reported values $Y_{it}$. The points are slightly jittered to show the
density of samples. The gray dashed line traces $X_{it}=Y_{it}$. Larger
reported counts often correspond to smaller estimated true counts,
possibly because the same subjects also reported very low counts at
other timepoints.

Estimates of $\theta_\mathrm{disp}$ are similar for all BDP models with
heaping, suggesting that dispersion or misremembering carries
information that is distinct from heaping or rounding in the data. The
regime parameters $\gamma_0,\ldots,\gamma_3$ are similar for all the
BDP heaping models, but likely not comparable to the WH08 model, as the
heaping mechanism is different. Estimates of the regime parameters can
be interpreted by transforming them into their regime transition
midpoints $-(\gamma_1, \gamma_2, \gamma_3)/\gamma_0$. For example, the
posterior mean estimates for the ``heaping'' model indicate that the
``no heaping'' regime dominates when the true count is between 0 and
$-\gamma_1/\gamma_0=10.7$ (posterior mean), and heaping to multiples of
50 dominates when the true count is greater than $-\gamma_3/\gamma
_0=16.2$. Between these values, heaping to multiples of 5 or 10
dominates. Estimates of $\gamma_1,\gamma_2,\gamma_3$ exhibit fairly
large posterior variance, and posterior intervals for $\gamma_1$ and
$\gamma_2$ show substantial overlap. This indicates that there is not
strong evidence of heaping to multiples of 5 and 10 in the data;
rather, small counts exhibit little heaping, and large counts show
strong heaping to multiples of~50.

We find that there is no significant difference in heaping by gender
under our model: the gender-specific effect $\omega$ in the last model
is not significantly different from zero. This finding is in contrast
to those of other researchers who see a strong effect of gender on
reporting of sexual behaviors [\citet{Wiederman1997Truth}]. One of the
goals of the CLEAR study was to show that educational intervention for
HIV-positive youth could reduce risky behaviors. While heaping behavior
may differ with respect to gender among subjects in the CLEAR study,
the small number of reported counts per subject does not permit us to
detect such a difference under the BDP heaping model. The intervention
tended to reduce true counts, and $\Pr(\alpha_\mathrm{intv}<0)>0.95$ for
every model.

We report two goodness-of-fit measures. The first is deviance
information criterion (DIC), computed by conditioning on posterior
samples of the parameters that directly affect the outcome $Y_{it}$.
For the ``no heaping'' model, these parameters are $\balpha$ and
$\bbeta
$; for the WH08 model, the $X_{it}$'s and $\bgamma$; for the
``dispersion-only'' model, the $X_{it}$'s and $\theta_\mathrm{disp}$; for
the ``heaping'' model, the $X_{it}$'s, $\theta_\mathrm{disp}$,
$\theta
_\mathrm{heap}$ and $\bgamma$; for the ``subject-specific heaping''
model, the $X_{it}$'s, $\theta_\mathrm{disp}$, $\bgamma$ and $\sigma
^2_\xi
$; and for the ``subject-specific heaping${}+{}$gender'' model, the
$X_{it}$'s, $\theta_\mathrm{disp}$, $\bgamma$, $\sigma^2_\xi$ and
$\omega
$. The second goodness-of-fit measure is the sum of squared mean
prediction errors, $\operatorname{SSPE} = \sum_{i=1}^n \sum_{t=1}^{n_i} (Y_{it}
- \hat{Y}_{it})^2$, where $\hat{Y}_{it}$ is the mean posterior
predictive value of $Y_{it}$, calculated by conditioning on the same
parameters as used to calculate the DIC. The \citet{Wang2008Modeling}
model is unique because $Y_{it}|X_{it}$ depends only on the four
rounding regimes parameters $\bgamma$, so the DIC is low, and the
heaping models all show similar DIC. The SSPE tells a different story:
the dispersion-only model shows the worst fit, and the BDP heaping
models outperform the WH08 model. These goodness-of-fit measures should
be interpreted carefully since the WH08 and BDP heaping models have a
somewhat different structure.

\begin{figure}[b]

\includegraphics{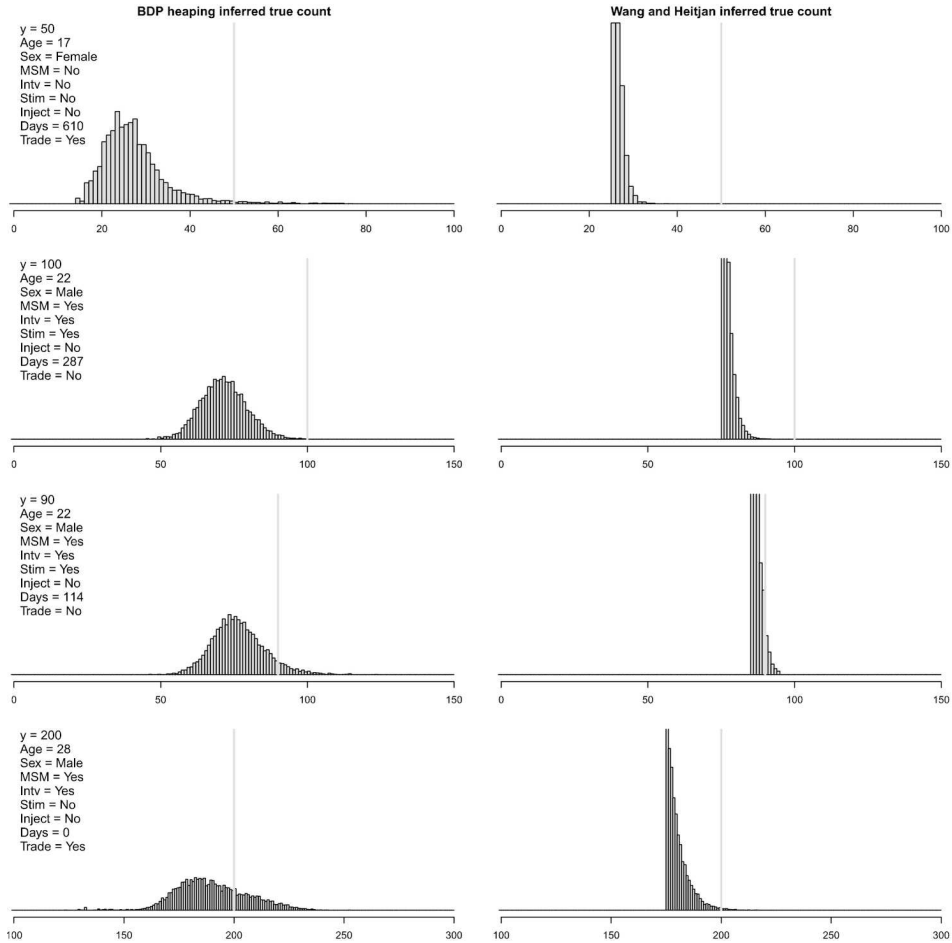}

\caption{Marginal posterior distributions of true counts $X_{it}$ for
individual subjects under the BDP heaping model with subject-specific
heaping parameters and the model of \citet{Wang2008Modeling}. The
subject- and timepoint-specific covariate values are listed with each
plot. A gray vertical line denotes the reported count $Y_{it}=y$. Not
all inferred true count distributions are centered at the reported
count. Moreover, the inferred true counts become more dispersed as the
reported count increases. The \citet{Wang2008Modeling} model does not
allow responses beyond the nearest heaping point and effectively puts a
uniform prior distribution on responses that fall within this window.
This results in inferred true counts whose posterior distribution is a
truncated version of the predictive distribution of $X_{it}$.}
\label{fig:postexample}
\end{figure}

The proportional odds model for different heaping regimes (rounding to
5, 10 and 50) introduced by WH08 proves to be an essential ingredient
in our analysis. The apparent heaping pattern observed in the CLEAR
counts of sex partners suggests that heaping to multiples of 50 happens
often as counts become larger than 30 or 40. We find that heaping
models that required rounding to multiples of~5, even for large counts,
provide a very poor fit (results not shown). However, in our analyses,
the model of WH08 has a serious drawback: when only one heaping regime
is in effect, it places a nearly uniform distribution on the true
count. The inferred true count distribution is proportional to the
product of this uniform distribution and the posterior predictive
distribution of the true count.
Figure~\ref{fig:postexample} illustrates the problem for specific subjects.
Both the WH08 model and the subject-specific BDP heaping model have
similar predictive distributions $f(x|\balpha,\bbeta)$ for the latent
true count $x$, and in both cases only the $v_3$ regime (rounding to
the nearest multiple of 50) is in effect. But the rounding model of
WH08 assumes that rounding is always to the \emph{nearest} grid point,
so, for example, a reported value of $y=200$ means that $x\in\{
175,\ldots,225\}$ with probability one. The heaping distribution
$g(y=200|x,\btheta,\bgamma)$ implicitly places a nearly uniform
distribution on this set, so the inferred posterior distribution of the
true count $x$ is a truncated version of $f(x|\balpha,\bbeta)$. In
contrast, the BDP heaping model provides a reporting distribution
$g(y=200|x,\bgamma,\btheta)$ that has support on all of $\mathbb{N}$
and preferentially places more mass on those $x$ that are most likely
to deliver the reported count $y$. In settings where the true counts
themselves might be the objects of inference, we believe the BDP
heaping model provides more realistic and useful estimates.


\section{Discussion}

In this paper we have illustrated how researchers can infer the
posterior distribution of true integer counts from reported counts
using a general BDP reporting distribution within a hierarchical
modeling framework. Our most substantial innovation is the novel
reporting distribution $g(y|x,\btheta)$ based on the BDP with specially
defined jumping rates that make the Markov chain attracted to heaping
grid points. Use of simple linear BDPs to model over-dispersion or
reporting error has been proposed before [\citet
{Grunwald2011Statistical,Lee2011Using}]. However, we have substantially
expanded the possibilities for general birth--death models of reporting
error to explicitly incorporate both over-dispersion and heaping, while
providing a computational method to evaluate likelihoods and sample
from the posterior distribution of the true counts. This approach has
the benefit of providing a sophisticated and highly configurable family
of reporting distributions indexed by the true count and just a few
unknown parameters $\btheta$ and $\bgamma$.

Statisticians may understandably be wary of parametric assumptions\break
about the way study participants report data. However, applied and
methodological research in public health offers some clues into
reporting mechanisms. Researchers in this field often address the
problem of reporting error in surveys related to sexuality and other
taboo topics [\citet{Schaeffer2000Asking}]. \citet{Wang2008Modeling}
discuss validation of reported counts of cigarettes smoked by measuring
tobacco products in the blood. In related work, \citet{Wang2012Truth}
compare instantaneous and retrospective self-reports of cigarette
consumption under a similar model for heaping. Other survey methods are
possible, including using diary-like surveys or repeated questionnaires
to assess reporting error. Studies like these can provide useful
information about the parameters $\btheta$ and $\bgamma$ in our BDP
heaping model. Armed with prior information about rounding
propensities, perhaps stratified by personal attributes such as gender,
age or sexual orientation, public health researchers could proceed with
a Bayesian analysis similar to the one outlined in this paper to
jointly estimate true counts and regression parameters. Designing a
model that accommodates various assumptions about both the mechanism
generating the true counts and the cognitive processes that give rise
to the reported counts can be challenging. The BDP model for heaped
counts presented in this paper is one promising step in this direction.



\begin{appendix}\label{app}
\section{Numerical evaluation of reporting probabilities}
\label{app:laplace}

We efficiently find the transition probabilities $P_{ab}(t)$ by first
applying the Laplace transform to both sides of the forward equations
[\citet{Karlin1957Differential,Murphy1975Some}]. This turns the infinite
system of differential equations \eqref{eq:odes} into a recurrence
relation whose solution yields an expression for the Laplace transform
of the transition probability $P_{ab}(t)$. To illustrate, let the
Laplace transform $h_{ab}(s)$ of the transition probability $P_{ab}(t)$ be
%
\begin{equation}
h_{ab}(s) = \int_0^\infty
e^{-st}P_{ab}(t)\,\mathrm{d} t.
\end{equation}
Then differentiating $h_{ab}(s)$ with respect to $t$ and setting
$a=b=0$, \eqref{eq:odes} becomes
%
\begin{eqnarray}\qquad
\label{eq:recur0} %
s h_{00}(s) - P_{00}(0) &=&
\mu_1 h_{01}(s) - \lambda_0
h_{00}(s)\quad \mbox{and}
\nonumber
\\[-8pt]
\\[-8pt]
\nonumber
s h_{0b}(s) - P_{0,b}(0) &=& \lambda_{b-1}h_{0,b-1}(s)
+ \mu_{b+1} h_{0,b+1}(s) - (\lambda_b +
\mu_b)h_{0b}(s)
\end{eqnarray}
for $b\geq1$. Rearranging \eqref{eq:recur0}, we find the recurrence
%
\begin{eqnarray}\label{eq:recur2}
h_{00}(s) &=& \frac{1}{s + \lambda_0 - \mu_1 (
{h_{01}(s)}/{h_{00}(s)} ) }\quad \mbox{and}
\nonumber
\\[-8pt]
\\[-8pt]
\nonumber
\frac{h_{0b}(s)}{h_{0,b-1}(s)} &= &\frac{\lambda_{b-1}}{s + \mu_b +
\lambda_b - \mu_{b+1} ({h_{0,b+1}(s)}/{h_{0,b}(s)} )}. %
\end{eqnarray}
From this recurrence, we arrive at the well-known continued fraction
representation for $h_{00}(s)$,
%
\begin{equation}
h_{00}(s) = \frac{1}{s+\lambda_0 - {\lambda_0 \mu
_1}/{(s+\lambda
_1+\mu_1 - {\lambda_1 \mu_2}/{(s+\lambda_2+\mu_2 - \cdots))}}} \label{eq:cfrac1}
\end{equation}
[see \citet{Murphy1975Some,Crawford2012Transition} for further details].
This is the Laplace transform of the transition probability
$P_{00}(t)$. From~\eqref{eq:cfrac1}, we can derive similar continued
fraction representations for $h_{ab}(s)$ for any $\CTMC(0)=a$ and
$\CTMC
(t)=b$. These expressions are given in the supplemental material [\citet
{Crawford2015HeapingSupplement}]. \citet{Crawford2012Transition}
present a numerical method for inverting transforms \eqref{eq:cfrac1}
to compute the transition probabilities in any general BDP with
arbitrary jumping rates $\{\lambda_k\}_{k=0}^\infty$ and $\{\mu_k\}
_{k=1}^\infty$. The supplementary material of \citet
{Crawford2014Estimation} shows how numerical error is controlled in the
computation. Section~\ref{app:X} of this appendix gives an
approximation to the reporting distribution that is useful for sampling.


\section{Approximation of reporting probabilities}
\label{app:X}

In this appendix we derive an approximation to the conditional
distribution of the reported count given the true count,
$Y_{it}|X_{it}$. The full conditional distribution of the $i$th
subject's true count $X_{it}$ at timepoint $j$ is
%
\begin{eqnarray}
&&\Pr(X_{it}=x\mid Y_{it},\Z_i,\W_{it},
\balpha,\bbeta_i,\btheta)\nonumber \\
&&\qquad\propto\Pr(Y_{it}\mid
X_{it}=x,\btheta) \Pr(X_{it}=x\mid\W_{it},\Z
_i,\balpha,\bbeta_i)
\nonumber
\\[-8pt]
\\[-8pt]
\nonumber
&&\qquad= P_{x,Y_{it}}(\btheta) \frac{\eta_{it}^x e^{-\eta_{it}}}{x!}
\\
&&\qquad= g(y|x,\btheta) f(x|\eta_{it}) ,\nonumber
\end{eqnarray}
where $\eta_{it} = \exp(\W_{it} \balpha+ \Z_{it} \bbeta_i)$ and
$P_{xy}(\btheta)=g(y|x,\btheta)$ is the general BDP transition
probability under the model described in Section~\ref{sec:bdmodel}.
Under a Metroplis--Hastings scheme, we need to propose a new value of
$X_{it}$ efficiently; we approximate the density $P_{xy}(\btheta)$ as
normal. Let $\theta_\mathrm{heap}=0$ and $\theta_\mathrm{disp}>0$. Then
this simplified BDP has birth and death rates
%
\begin{equation}
\lambda_k= \theta_\mathrm{disp} + \theta_\mathrm{disp}
k \quad\mbox{and}\quad \mu_k=\theta_\mathrm{disp} k.
\end{equation}
This is a linear process with immigration that has an asymptotically
normal distribution. Similar to Section~\ref{sec:bdps}, let $\CTMC(t)$
be a BDP starting at $\CTMC(0)=a$. Following \citet{Lange2010Applied},
we form the probability generating function (PGF)
%
\begin{equation}
H(s,t) = \sum_{b=0}^\infty s^b
P_{ab}(t) ,
\end{equation}
where $s$ is a ``dummy'' variable and $P_{ab}(t)=\Pr(\CTMC(t)=b \mid
\CTMC(0)=a)$ is the transition probability. Although $H(s,t)$ has a
closed-form solution that can be inverted to obtain the $P_{ab}(t)$ in
analytic form, the details are somewhat complicated, and we only
require a normal approximation to this density. The mean $m_a(t)=\E
(\CTMC(t) \mid\CTMC(0)=a)$ is given by
%
\begin{equation}
\frac{\partial H(s,t)}{\partial s}\bigg|_{s=1} = \sum_{b=0}^\infty
jP_{ab}(t) = \E \bigl[\CTMC(t) \bigr] = m_a(t),
\end{equation}
and likewise the second factorial moment $e_a(t)$ is given by
%
\begin{equation}
\frac{\partial^2 H(s,t)}{\partial s^2}\bigg|_{s=1} = \sum_{b=1}^\infty
b(b-1)P_{ab}(t) = \E \bigl[\CTMC(t)^2 \bigr] - \E \bigl[
\CTMC(t) \bigr] = e_a(t),
\end{equation}
where the expectations are conditional on the process beginning in
state $\CTMC(0)=a$. This suggests that we can determine the mean and
variance of $\CTMC(t) \mid\{\CTMC(0)=a\}$ by finding the partial
derivatives of $H$ with respect to the dummy variable $s$. To derive
these quantities, we form a partial differential equation for the
solution of the PGF
%
\begin{equation}
\frac{\partial H(s,t)}{\partial t} = \theta_\mathrm{disp} \biggl[ (s-1)^2
\frac{\partial H(s,t)}{\partial s} + (s-1) H(s,t) \biggr]. \label{eq:pgf}
\end{equation}
See \citet{Lange2010Applied}, \citet{Bailey1964Elements} and \citet
{Renshaw2011Stochastic} for the details of deriving this generating
function. Now, the time-derivative of the mean falls out as
%
\begin{equation}
\frac{\mathrm{d} m_a(t)}{\mathrm{d} t} = \frac{\partial^2
H(s,t)}{\partial t\,\partial
s}\bigg|_{s=1} = \theta_\mathrm{disp}
,
\end{equation}
and the time-derivative of the second factorial moment is
%
\begin{equation}
\frac{\mathrm{d} e_a(t)}{\mathrm{d} t} =\frac{\partial^3
H(s,t)}{\partial t\,\partial^2
s}\bigg|_{s=1} = 4 \theta_\mathrm{disp}
(a+\theta_\mathrm{disp} t).
\end{equation}
Solving these differential equations with the initial conditions
$m_a(0)=a$ and $e_i(0)=a^2 - a$ yields
%
\begin{equation}
m_a(t) = a + \theta_\mathrm{disp} t \quad\mbox{and}\quad
e_a(t) = a(a-1) + 4a\theta_\mathrm{disp} t + 2
\theta_\mathrm{disp}^2t^2.
\end{equation}
From these, we determine that
%
\begin{eqnarray}
\E \bigl[\CTMC(t)\mid\CTMC(0)=a \bigr] &=& a + \theta_\mathrm{disp}
t \quad\mbox{and}
\nonumber
\\[-8pt]
\\[-8pt]
\nonumber
\operatorname{Var} \bigl[\CTMC(t)\mid\CTMC(0)=a \bigr] &=& (2a+1)\theta
_\mathrm{disp} t + \theta_\mathrm{disp}^2
t^2 , %
\end{eqnarray}
where the second line arises because $\operatorname{Var}[\CTMC(t) \mid\CTMC
(0)=i] = e_a(t) + m_a(t) - m_a(t)^2$. Therefore, a reasonable
approximation to the probability mass function of $\CTMC(t) \mid\{
\CTMC
(0)=a\}$ is the normal distribution with the mean and variance above.
This approximation serves as an effective proposal within a
Metropolis--Hastings accept/reject step.
\end{appendix}

\section*{Acknowledgments}
We thank Kenneth Lange, Janet Sinsheimer and\break Gabriela Cybis for
thoughtful comments. We also acknowledge Robert D. Bjornson and
Nicholas J. Carriero for providing cluster computing resources at Yale.


\begin{supplement}[id=suppA]
\stitle{Supplemental article}
\slink[doi]{10.1214/15-AOAS809SUPP} 
\sdatatype{.pdf}
\sfilename{aoas809\_supp.pdf}
\sdescription{We provide a derivation of the Laplace transform of
transition probabilities for a general BDP, the full posterior
distribution and an outline of Monte Carlo sampling procedures for
unknown parameters.}
\end{supplement}



%
%


\printaddresses
\end{document}